# Nanoscale ferroelectric manipulation of magnetic flux quanta

Arnaud Crassous[1], Rozenn Bernard[1], Stéphane Fusil[1,2], Karim Bouzehouane[1], D. Le Bourdais[1], Shaïma Enouz-Vedrenne[3], Javier Briatico[1], Manuel Bibes[1], Agnès Barthélémy[1] and Javier E. Villegas[1,*]

[1] *Unité Mixte de Physique CNRS/Thales, 1 ave. A. Fresnel, Campus de l'Ecole Polytechnique, 91767 Palaiseau, and Université Paris Sud 11, 91405 Orsay, France*

[2] *Université d'Evry-Val d'Essonne, Boulevard François Mitterrand, 91025 Evry, France*

[3] *Thales Research and Technology, 1 Avenue Augustin Fresnel, Campus de l'Ecole Polytechnique, 91767 Palaiseau, France*

Using heterostructures that combine a large-polarization ferroelectric ($BiFeO_3$) and a high-temperature superconductor ($YBa_2Cu_3O_{7-\delta}$), we demonstrate the modulation of the superconducting condensate at the nanoscale via ferroelectric field effects. Through this mechanism, a nanoscale pattern of normal regions that mimics the ferroelectric domain structure can be created in the superconductor. This yields an energy landscape for magnetic flux quanta and, in turn, couples the local ferroelectric polarization to the local magnetic induction. We show that this form of magnetoelectric coupling, together with the possibility to reversibly design the ferroelectric domain structure, allows the electrostatic manipulation of magnetic flux quanta.

[*] email: javier.villegas@thalesgroup.com

Many complex oxides are strongly correlated electrons systems in which slight variations of the charge carrier density lead to dramatic changes of the physical properties[1]. In addition to its fundamental interest, this characteristic offers significant potential for novel technological applications, as it expands the possibilities of the electrostatic tuning of the carrier concentration ("field-effect doping")[2] from the simple manipulation of charge currents –as in classical field-effect transistors– to the control of functional properties such as ferromagnetism[3,4] or superconductivity[5-13]. Oxide superconductors are an archetypal example. Their critical temperature $T_C$ can be changed via the application of an electric field, using a dielectric gate as in conventional transistors[5-7,10,11,13], or via ferroelectric field effects[8,9,12]. In the latter case, the carrier density modulation is induced by switching the polarization of a top ferroelectric layer upon the application of a voltage pulse. This effect allows changing the $T_C$ in a nonvolatile and reversible way. A complete switching of superconductivity has been achieved in this fashion in heterostructures combining a low-carrier-density superconductor and a ferroelectric at very low temperatures T~ 200 mK[8], while for high-$T_C$ films, a relatively weaker variation of the critical temperature $\Delta T_C$ ~ 7 K has been obtained[9,10].

Based on the possibility of "writing" ferroelectric domains using atomic force microcopy techniques[14], one of the groundbreaking prospects of the ferroelectric field effects described above is their potential for the fabrication of reprogrammable circuits that might exploit unique superconducting properties[8] –e.g. magnetic flux quantization and Josephson coupling–. However, that proposal has remained as such since, in fact, the ferroelectric modulation of superconductivity has been demonstrated only over relatively long length scales (as compared to the relevant superconducting lengths): in the best case, gating areas with lateral sizes in the tenths-of-microns have been used[8]. Fundamentally, the outstanding question is whether strong local electrostatic doping effects can be produced over areas much smaller than that or if, on the contrary, the $\Delta T_C$ contrast between neighboring oppositely-doped regions fades out as their area is further scaled down. The question arises since, due to the relatively high carrier concentration in these materials (as compared to semiconductors)[1], one could expect the local doping to average out to the background level when the gating area is significantly reduced. In addition, the superconducting proximity-effect[15] could quench the $\Delta T_C$ contrast for gate dimensions comparable to the superconducting condensate decay-length across superconducting-normal interfaces.

In this Letter, we experimentally demonstrate ferroelectric field-effect doping of superconductors at the nanoscale. We show that this effect can be used to induce a strong, remanent, tunable nanometric spatial modulation of the $T_C$, whose geometry mimics that of the ferroelectric domain structure. This creates an energy landscape for magnetic flux quanta, in which the energy wells form in regions where the $T_C$ is depressed due to the local ferroelectric polarization. As a result, the field-effect doping ultimately causes a new form of magnetoelectric coupling, which is between the local polarization in the ferroelectric and the local magnetic induction in the superconductor. Such coupling is strong enough to govern flux dynamics, and prominently shows up in the macroscopic magneto-transport. The key to this realization is that, due to the materials choice and optimization, we obtain upon ferroelectric switching a uniquely strong *persistent* $\Delta T_C$, up to ~30 K, the largest known in cuprates. Because we can reversibly design and modify the ferroelectric domain structure at will to create artificial nanoscale patterns, the effects we demonstrate open the door to a plethora of studies and applications. These include, in addition to tunable superconducting circuits (such as Josephson junctions arrays, nano-SQUIDs, etc), reconfigurable energy landscapes to control magnetic flux quanta in fluxtronics devices[16-20], as well as studies on nanoscale confinement, dimensionality and proximity effects in high-temperature superconductors. In addition, the nanoscale field-effect doping demonstrated here may be applied to other strongly correlated electrons systems with relatively high carrier density (e.g. oxide ferromagnets) for analogous nanoscale studies and applications.

In order to produce strong and nonvolatile field effects, the strategy is to combine an ultra-thin superconducting film and a ferroelectric material with a large remanent polarization[1]. Here we use the archetypal high-$T_C$ superconductor YBa$_2$Cu$_3$O$_{7-\delta}$ (YBCO) and BiFeO$_3$ (BFO), a ferroelectric[21] with ABO$_3$ perovskite structure and one of the largest known bulk ferroelectric polarizations [up to 100μC.cm$^{-2}$ along the (111) direction[22]]. C-axis YBCO/BFO bilayers (BFO at the top) were epitaxially grown *in situ* on (001) oriented SrTiO$_3$ (STO) using Pulsed Laser Deposition (PLD). A buffer layer of PrBa$_2$Cu$_3$O$_7$ (PBCO, a semiconductor isostructural to YBCO) was grown on the STO substrate prior to YBCO deposition in order to optimize $T_C$. However, PBCO is not expected to play any role on the field effects discussed here. X-Ray structural characterization (not shown) proved that BFO can be combined with YBCO in high structural quality heterostructures. High-resolution transmission electron microscopy [Fig. 1 (b)] confirms c-axis growth and shows a coherent interface without misfit dislocations between BFO and YBCO. The heterostructures used for the electrical transport experiments discussed below are BFO(30 nm)/YBCO($t_{YBCO}$)/PBCO($t_{PBCO}$)//STO heterostructures (u.c.=unit cell), with the YBCO thickness $t_{YBCO}$ ranging from 3 to 6 u.c., and $t_{PBCO}$ either 2 or 4 u.c.

A multi-probe bridge for transport experiments was optically lithographed and ion etched onto the samples (see Fig. 1 (a) ; two of the voltage probes –V$^+$ and V$^-$– and the flow direction of the in-plane current *J* injected for electrical transport experiments are indicated). An atomic force microscope (AFM) operating in piezoresponse mode (PFM) was used to image the ferroelectric polarization of the BFO layer at room temperature. In addition to imaging, this set-up can be used to artificially create ferroelectric domains patterns[14,23]. For this, a dc voltage $V_{DC}$ is applied between the conductive tip of the AFM and the YBCO layer.

If the electric field generated across the BFO is above the coercive field, it will locally switch the ferroelectric polarization, which is typically achieved by the application of $|V_{DC}|$~3 V in the samples investigated here [see Fig. 1 (c)].

In the virgin state, the out-of-plane component of the BFO polarization is homogeneous and points towards the YBCO layer (hereafter "down" polarization). This is shown in Fig. 1 (d), which displays the PFM image of the bridge prior to any manipulation. To reverse the ferroelectric polarization $V_{DC}$ ~5 V was applied as the AFM tip was scanned over the selected area. The PFM image in Fig. 1 (e) is taken after the polarization was reversed within the area between the two voltage probes. The dark contrast indicates a phase change of 180º, which implies that the out-of-plane component of the polarization points outwards from the YBCO layer (hereafter "up" polarization). The sample surface topography is unaffected by repeated poling and reading.

Since YBCO is a hole-doped superconductor[24], when the BFO polarization points "up" an accumulation of charge carriers ("doping") and therefore an increase of $T_C$ is expected. Conversely, if the polarization points "down", a charge carriers depletion and a depression of $T_C$ can be anticipated. This is demonstrated in Fig. 2 (a), which displays the resistance *vs.* temperature *R(T)* for two different polarization states for a sample with $t_{YBCO}$=3 u.c.. The blue curve corresponds to the case in which the ferroelectric polarization in the area between the voltage probes is the virgin state ("down" polarization, depleted state), and the red one to the case in which the polarization is "up" (doped state). A large difference between the critical temperatures $\Delta T_C$ is observed.

Fig. 2 (b) shows $T_C$ *vs.* the charge carrier density *n* for samples with different $t_{YBCO}$ (see legend). For each sample, *n* and $T_C$ were measured in the same sample area (i.e. between a fixed pair of voltage probes) i) for a "down" polarization (depleted state, hollow symbols), and ii) after the polarization has been switched "up" at room temperature (doped state, solid symbols). $T_C$ is defined (see inset) with the criterion $R(T_C)=0.1 R_N$ (with $R_N$ the normal-state resistance at the onset of the transition). $n$ was obtained from Hall-effect experiments in the normal-state (just above the superconducting transition onset). In all cases, the $\Delta T_C$ induced upon ferroelectric switching is accompanied by a consistent change $\Delta n$. $T_C(n)$ for bulk YBCO is plotted as a reference (dashed line).

The maximum $\Delta T_C \sim 30$ K ($\Delta n = 5.2\ 10^{20}$ cm$^{-3}$) corresponds to the thinnest $t_{YBCO}=3$ u.c. This is much in excess of previous realizations of ferroelectric modulation of superconductivity[8] in which $\Delta T_C \sim 7$ K. Note in particular [see inset in Fig. 2 (b)] that reversing the polarization causes a complete switching of superconductivity: at temperatures ~ 35 K, the sample is either in the normal or in the superconducting state depending on the ferroelectric polarization. $\Delta T_C$ is weaker for thicker $t_{YBCO}$, with the smallest $\Delta T_C \sim 7$ K corresponding to $t_{YBCO} = 6$ u.c. and $\Delta n = 6\ 10^{19}$ cm$^{-3}$. From the data in Fig. 2 (b), the largest sheet carrier density variation between the depleted and doped state is $\Delta n_{sh}^{exp} = \Delta n \cdot t_{YBCO} \sim 1.9 \cdot 10^{14}$ holes·cm$^{-2}$, for $t_{YBCO}=3$ u.c. This is only 25% of $\Delta n_{sh}^{theo} = 2P/e \sim 8 \cdot 10^{14}$ holes·cm$^{-2}$ expected from the polarization of the BFO along the (001) direction $P \sim 65$ µC·cm$^{-2}$ (Ref. 25). The disagreement is probably due to the fact that the latter estimate assumes that the charge depletion/accumulation within the YBCO layer is homogeneous along the c-axis direction, which is indeed unlikely[11] provided that the YBCO is thicker than the Thomas-Fermi screening length ~1 u.c[7]. Furthermore, screening due to the presence of interface trap states cannot be excluded, which would also reduce the effective interface charge density and thus the field effects. Nevertheless, BFO can induce a relatively large variation of the sheet carrier density as compared to other ferroelectrics, in particular up to ~160% of that produced by Pr(Zr,Ti)O$_3$, for which $\Delta n_{sh}^{exp} \sim 1.2 \cdot 10^{14}$ holes·cm$^{-2}$ (Ref. 8). This is key in producing the significantly larger $\Delta T_C$ obtained here.

We verified that the field effects described above are persistent at least for 8 days, and that (aside from minor training effects) the $T_C$ shift is essentially reversible, as expected[8]. This was done by performing cycles in which the BFO polarization was alternatively switched up/down and the $R(T)$ subsequently measured.

We show in what follows that the strong field effects above can be used to produce a nanoscale modulation of the superconducting condensate. To this end, we created periodic arrays of ferroelectric domains. This was done via a two step process. First, the area between the voltage probes was homogeneously poled "up" ("doped" state) using the AFM as described previously. Subsequently, -8.5 V pulses (20 ms in duration) were locally applied between the YBCO layer and the AFM tip, periodically in space along the $x$ and $y$ directions. This locally reversed the polarization, creating a periodic array of nano-domains ("dots" with a diameters $\varnothing \sim 30\text{-}80$ nm) in which the polarization points "down". PFM images of the obtained arrays are displayed in Fig. 3 (a) (array A) and 3 (b) (array B). In both cases, the array unit cell or *plaquette* is a parallelogram with sides $a \neq b$. After definition of those ferroelectric patterns, the $T_C$ is comparable to the case in which the ferroelectric has a homogeneous "up" polarization.

Evidence for the nanoscale modulation of superconductivity induced by the ferroelectric arrays is obtained from the mixed-state magneto-resistance. The black curves in Fig. 3 (c) and (d) show to the resistance versus $H$ (applied perpendicular to the film plane), for the arrays A and B respectively. In both cases, the in-plane current $J$ is applied parallel to the array *plaquette* base, $b$. For the array A [Fig. 3 (c)], two local minima appear symmetrically around $H = 0$ at the fields $|H_1| = 960 \pm 40$ Oe. This behaviour must be compared to the monotonous magneto-resistance exhibited by the red curve, which was measured after the array A was "erased" by subsequently switching "down" and "up" the BFO polarization using the AFM, which results in an homogeneous "up" state [as in Fig. 1 (e)]. For array B [Fig. 3 (d)], two local minima are observed at fields $|H_1|$ = 1400±50 Oe and, in addition, a pair of weaker features can be seen at $|H_2| = 2|H_1|$ = 2800±50 Oe.

The magneto-resistance minima observed in the presence of the ferroelectric arrays are the well-known fingerprint of periodic pinning of flux quanta[16-18,26-28]. The matching fields are as expected for the present array geometries. For the array A, the minima appear at fields $|H_1|$ for which the distance between flux quanta $d = 1.075(\phi_0/\mu_0 H_1)^{1/2}$ =161±3 nm matches the side $b$=157±3 nm of the array *plaquette*. I.e., the flux pinning enhancement appears due to the commensurability between the triangular flux lattice and the pinning potential in the direction parallel to the injected current. This is as expected[28] for arrays in which the distance between pinning sites along the current direction is the shortest ($b < a$). For the array B, the matching field is $|H_1|$ = 1400±50 Oe. This agrees with $H_\phi = \phi_0/S$ =1320±60 Oe, with $S = a\sin(\alpha)\times b$ the *plaquette* area, $a$=113±3 nm, $b$=143±3 nm and the base angle $\alpha$ =106.0°±0.5°. That is, for the array B, the matching occurs in the presence of an integer number of flux quantum per pinning site in the array. This implies that the flux lattice looses its triangular geometry and matches that of the ferroelectric array at the fields $|H_1|$, $2|H_1|$, etc. This type of commensurability is as expected when the distance between pinning sites along the current direction is the longest ($a < b$) (Ref. 28). The angular dependence of the commensurability effects is consistent with the flux pinning scenario. When $H$ is applied at increasing angles $\theta$ with respect to the c-axis [Figs. 3 (e)], the background magneto-resistance diminishes –as expected for anisotropic superconductors[29]– and the matching fields increase. As shown in Fig. 3 (f), these scale as $1/\cos(\theta)$. This implies that the matching phenomena solely depend on the component of the applied field perpendicular to the sample surface, as expected for commensurability effects in superconducting thin films with artificial periodic arrays[27]. The temperature and current dependences of these effects (not shown) are also as expected for periodic flux pinning in superconductors.

The above results imply that the geometry of the ferroelectric domain structure is transferred into the YBCO via the local modulation of the superconducting critical temperature. A sketch of the mechanism is shown in Fig. 3 (g). The electric field from "down" ("up") polarized nano-domains produces a local depletion (doping) of charge carriers and consequently a local depression (enhancement) of $T_C$. Thus, at temperatures between the maximum and minimum $T_C$, a periodic distribution of nanometric "dots" where superconductivity is suppressed (or strongly depressed) form in the YBCO film. This is strong enough to pin flux quanta, which gives rise to the commensurability effects characteristic of periodic flux pinning evidenced by the R(H) curves in Fig. 3.

In summary, we have demonstrated the electrostatic pinning of magnetic flux quanta. For this, we obtained a nanoscale modulation of high-temperature superconductivity via ferroelectric field effects. This creates an energy landscape for flux quanta, which couples the local ferroelectric polarization to the local magnetic induction in the superconductor. In addition to its fundamental interest, the possibility to manipulate flux quanta by designing the structure of ferroelectric domains is relevant in view of applications. These include fluxtronic devices based on the controlled motion of flux quanta[16-20], for which the ferroelectric approach has two key advantages i) the pinning potential geometry is reconfigurable and ii) its lateral length scale is much shorter than typically achieved by lithography techniques in high-$T_c$ films[18], which allows the manipulation of much higher densities of flux quanta. Besides different types of reprogrammable nanometric superconducting circuits, the possibility we demonstrate to produce ferroelectric field-effect doping at the nanoscale in systems with relatively high charge carrier densities might be used for the fabrication of reconfigurable nano-devices based on other correlated oxides[30]. For example, spintronic devices using oxide ferromagnets[3], or quantum circuits based on semiconductors. Examples of these have been recently realized using the 2D electron gas formed at the interface between two band insulators ($SrTiO_3$ and $LaAlO_3$)[31], via a physical mechanism that seems to be specific to this particular system[32]. Contrary to this, the nanoscale ferroelectric field effects demonstrated here are general, and may be applied to any system in which the physical properties are strongly dependent upon the charge carrier density[1].

We acknowledge C. Visani, V. Garcia and N. Reyren for useful discussions, E. Jacquet for technical support for thin film deposition, and D. Sando for critical reading of this manuscript. This work was supported by the French ANR via "OXITRONICS", "Méloic" and "SUPERHYBRIDS-II" grants.

**Figure Captions**

**Fig. 1.** (a) AFM of the multi-probe measurement bridge (b) HRTEM of a BFO(30 nm)/YBCO(5 u.c.)//STO heterostructure. (c) Local out-of-plane piezoresponse phase hysteresis loop. (d) PFM phase image in the pristine state (prior to poling). (e) PFM phase image after the area in between the voltage probes has been poled "up". For a sample with $t_{YBCO}$=3 u.c..

**Fig. 2.** (a) R(T) (normalized to the resistance at T=150 K) of a heterostructure with $t_{YBCO}$=3 u.c. measured with J=1.7 kA·cm$^{-2}$ for two neighboring areas in which the BFO polarization points "down" (blue curve) and "up" (red curve). The inset shows schematics of the charge carrier depletion or accumulation induced in the YBCO by the ferroelectric polarization (white arrows). (b) $T_C$ vs. charge carrier density $n$ for samples with different $t_{YBCO}$ (see legend in u.c.), for the "depleted" (hollow symbols) and "doped" (solid symbols) states. The inset shows R(T) corresponding to a single are within a sample with $t_{YBCO}$=3 u.c. in the virgin state (blue) and after the polarization has been switched (red), measured with J=1.7 kA·cm$^{-2}$. The stars mark $T_C$.

**Fig. 3.** (a) PFM phase image of the ferroelectric nano-domain array **A**, defined in the BFO layer of a heterostructure with $t_{YBCO}$=3 u.c. The array *plaquette* is a parallelogram with $b$=157±3 nm, $a$=195±3 nm and base angle $\alpha$=104.0°±0.5°. (b) Same for array **B** in a heterostructure with $t_{YBCO}$=4 u.c., with $b$=143±3 nm, $a$=113±3 nm and $\alpha$=106.0±0.5± (c) Resistance (normalized to the zero-field one) *vs.* applied field at T=1.05$T_C$ and with injected current density J=8.7 kA·cm$^{-2}$ for a heterostructure with $t_{YBCO}$=3 u.c, in the case in which the periodic ferroelectric array A is present (black curve) and after this has been erased (red curve) at T=0.94$T_C$. (d) Resistance (normalized to the zero-field resistance) *vs.* field applied perpendicular to the film plane for a heterostructure with $t_{YBCO}$=4 u.c, at a T=0.99$T_C$ and with J=3.3 kA·cm$^{-2}$, in the case in which the periodic ferroelectric array B is present (black curve).

The vertical lines point out the matching fields. (**e**) $R(H)$ in the presence of the periodic array A, with $H$ applied at different angles $\theta$ =0º, 30º, 45º, 60º, 70º (top to bottom) with respect to the c-axis (see sketch). The vertical lines point out the matching fields for the different $\theta$. (**f**) Matching field $H_1$ as a function of $1/\cos(\theta)$ for the array A (squares) and B (circles). The straight line is the best linear fit. (**g**) Schematics of the periodic charge carrier depletion/accumulation and $T_C$ modulation induced by the structure of ferroelectric domains.

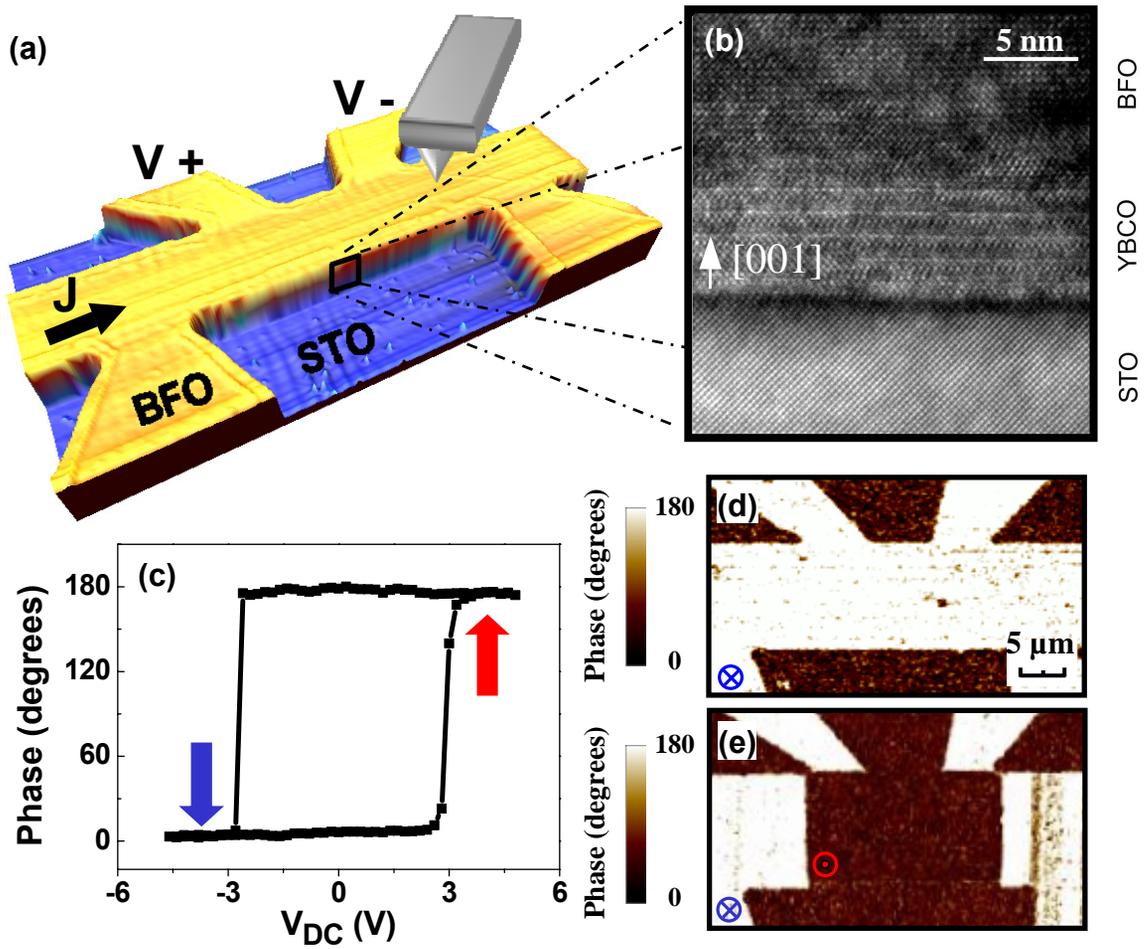

Figure 1

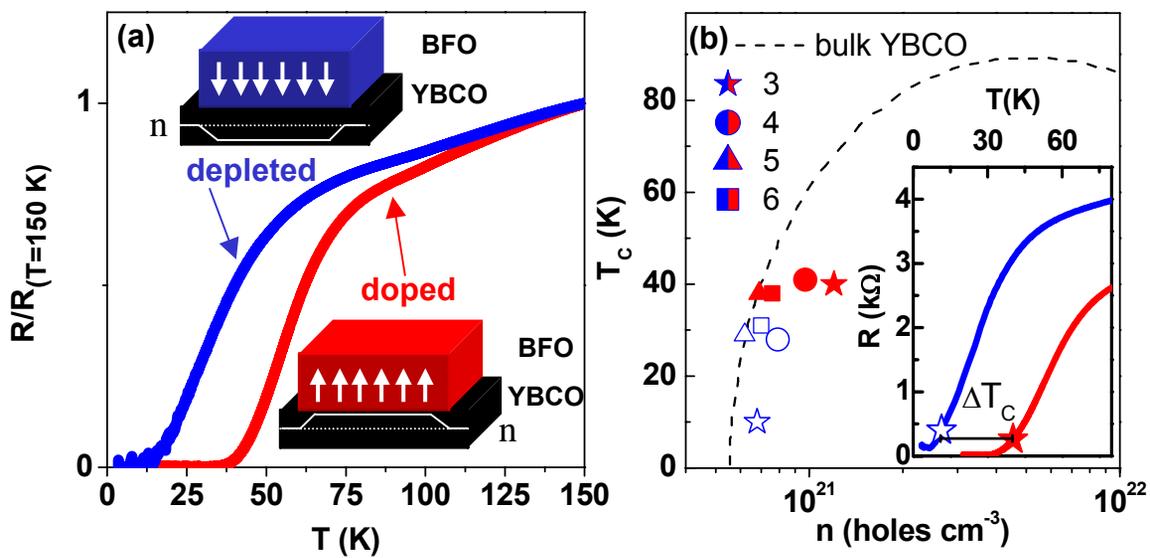

Figure 2

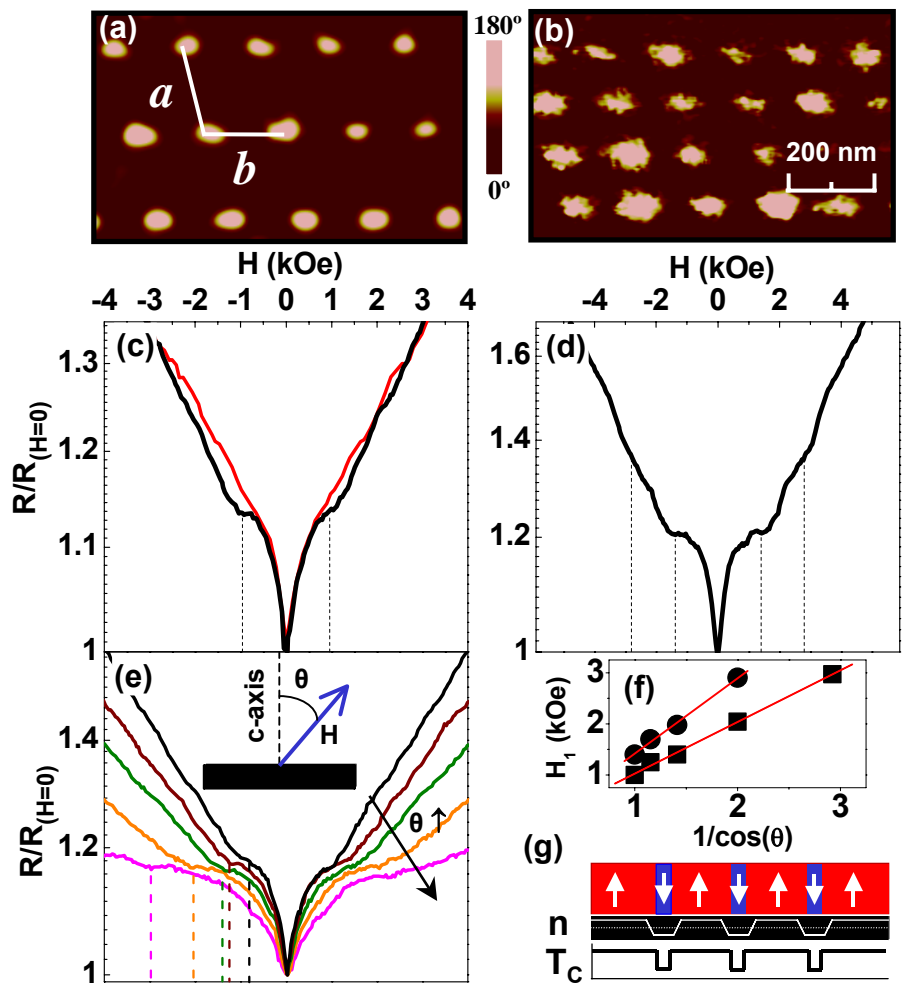

Figure 3